\documentclass[12pt]{article}
 \pdfoutput=1
 \usepackage{epsfig}
 \usepackage{color}
 \usepackage{amsmath}
 \usepackage{amssymb}
 \def\aprle{\buildrel < \over {_{\sim}}}

\begin{document}       

 \title{Relic antineutrino capture on $^{163}$Ho decaying nuclei}

 \author{Maurizio Lusignoli  and Marco Vignati \\\\
 {\normalsize
 Sapienza, Universit\`{a} di Roma, and INFN, Sezione di Roma}
\\
\normalsize
 Piazza A. Moro 2, I-00185 Roma, Italy 
} 

 \date{}          
 \maketitle

\begin{abstract}
The electron capture decay of the isotope $^{163}$Ho has been proposed since a long time as a candidate for measuring the electron neutrino mass and recently the interest on this idea has been renewed. In this letter we note that a direct observation of the cosmic antineutrino background could be made using a target made of this isotope. We further discuss the requirements for an experiment aiming to obtain this result. 
\end{abstract}

\section{Introduction}
For many years the possibility of direct observation of the cosmic  neutrino background (C$\nu$B), namely of those neutrinos that are relics of the Big Bang, has been considered and its terrible difficulty stressed.  We know that their average number density in the Universe should be 
$n_{\nu} \sim 55$ cm$^{-3}$ for neutrinos (or antineutrinos) of each flavour and that they decoupled at a temperature around 1 MeV. Since then, the Universe expanded  by about 10 orders of magnitude and therefore the average C$\nu$B momenta are today $\sim 10^{-4}$~eV.

Deviations of the spectrum in beta decays near to the end point due to a possible neutrino chemical potential in C$\nu$B were first suggested many years ago by Weinberg \cite{Weinberg:1962zz}, but present limits on the chemical potential from nucleosynthesis \cite{BBN} make this effect unobservable. It is the effect of the nonzero mass of the neutrino\footnote{As usual, we mean by this the mass of the mass eigenstate more strongly coupled to the electron neutrino state.} 
that could instead provide a hope, if its value is close to the present experimental bound. If this is the case, gravitational clustering could increase the number density $n_{\nu}$ by one or two orders of magnitude~\cite{Ringwald:2004np}. We recall that present limits are $\aprle 2$~eV from tritium decay  \cite{Lobashev:2003kt,Kraus:2004zw} and $\aprle 0.5$~eV from cosmology \cite{Fogli:2008ig,Thomas:2009ae}. 
 
It has been  recently proposed \cite{Cocco:2007za} to try to observe the process of capture of a neutrino in the  C$\nu$B by a $\beta$--decaying nucleus. In this case the electron in the final state has energy larger than the maximum energy of $\beta$--rays by  twice the value of the neutrino mass, and could therefore be distinguished, with a \textit{very} good resolution, if the neutrino mass is large  enough. An obvious candidate for the target is tritium, due to its small $Q$--value (18.6 keV)  and  good sensitivity to neutrino mass effects. 

The detection of antineutrinos in C$\nu$B could analogously be made using as target radioactive atoms that decay by electron capture (EC). This possibility has been examined in ref. \cite{Cocco:2009rh}, but apparently discarded as much less promising. In this letter, we want to deepen the examination in order to show that this is not correct, and that in fact the capture of antineutrinos in the  C$\nu$B by nuclei of $^{163}$Ho (the record element for low $Q$--value in EC decays) could be a valid alternative.

\section{Electron capture in $^{163}$Ho}
\label{theory1}
The energy spectrum of neutrinos produced in EC decays is given by a series of lines, each at an energy $Q-E_i$ (where $Q$ is the mass difference of the two atoms in their ground states and $E_i$ is the binding energy of the electron hole in the final atom). The decay process that we consider is:
\begin{equation}
^{163}{\rm Ho}  \to  \; ^{163}{\rm Dy}_i ^*+ \nu_e  \to  \; ^{163}{\rm Dy} + E_i + \nu_e \;.
\label{decay}
\end{equation} 
The EC decay rate can be expressed, following  \cite{Bambynek:1977zz}, as a sum 
over the possible captured levels: 
\begin{equation}
\label{ECrate}
\lambda_{EC} = {G_{\beta}^2 \over {4 \pi^2}} \; \sum_i n_i \, C_i\,  \beta_i^2 B_i\, (Q-E_i)[(Q-E_i)^2-m_{\nu}^2]^{1/2} \; .
\end{equation}  
In this equation $G_{\beta} = G_F \cos \theta_C$, $n_i$ is the fraction of occupancy of the i-th atomic shell, $C_i $ is the nuclear shape factor, $\beta_i$ is the Coulomb amplitude of the electron radial wave function (essentially, the modulus of the wave function at the origin) and $B_i$ is an atomic  correction for electron exchange and overlap. 
 The spin and parity of the nuclei involved in reaction (\ref {decay}) obey the relations 
 $\Delta J= 1,\;\Pi_f \Pi_i = +1$, and the transition is dubbed as allowed. The $Q$--value for this reaction is so small that only electrons from levels $M_1, M_2, N_1, N_2, O_1, O_2, P_1$ can be captured. In a very good approximation the nuclear shape factors $C_i$ are all equal in an allowed transition,  as it has been discussed in  ref.~\cite{Bambynek:1977zz}, and can be factored out from the sum in eq.(\ref{ECrate}). Different determinations of the  $Q$--value can be found in the literature \cite{NDS}. In this letter we will use values ranging from an optimistic 2.3~keV to a pessimistic 2.8~keV. 
 
The low Q-value of this transition prompted many years ago several proposals to use $^{163}$Ho decays to search for a signal of nonzero neutrino mass (as opposed to the antineutrino mass measured in tritium decays). The neutrino mass in fact affects the capture rates from different levels  \cite{Bennett:1981}, as in eq.(\ref{ECrate}),  and it modifies the spectra of inner brehmstrahlung photons  \cite{DeRujula:1981ti} and emitted electrons \cite{DeRujula:1982bq} near to their endpoints. Several experiments have been performed \cite{expHo},  obtaining upper bounds on the neutrino mass larger than 200 eV.  These measurements were mainly limited by the poor knowledge of complicated atomic corrections that can modify the emitted X--rays spectrum.

A more promising technique would be a calorimetric experiment, embedding the radioactive $^{163}$Ho source in a bolometer. This was suggested many years ago \cite{DeRujula:1982qt} and it is presently being developed \cite{Gatti:1997}  thanks to the huge improvements of the technique in terms of energy resolution. The advantage with respect to other techniques is that all the de--excitation energy is measured and does not remain partly trapped in invisible channels. The atomic levels have a finite (albeit often small) natural width, and therefore the lines have a Breit--Wigner resonance form, so that the spectrum of ``calorimetric'' energy $E_c$ should be given by\footnote{Some justifications concerning the neglect of interference terms and the absence of corrections due to final particles phase space have been given in  \protect\cite{DeRujula:1982qt}. }

\begin {eqnarray}
\label{E_c-distr}
{d \lambda_{EC}\over dE_c} &=& {G_{\beta}^2 \over {4 \pi^2}}(Q-E_c) \sqrt{(Q-E_c)^2-m_{\nu}^2} \;
\times \nonumber \\
&& \sum_i n_i  C_i \beta_i^2 B_i {\Gamma_i \over 2\,\pi}{1 \over (E_c-E_i)^2+\Gamma_i^2/4} \;. 
\end{eqnarray}

\begin{figure}[ht]
 \centering
  \includegraphics[width=0.7\textwidth ]{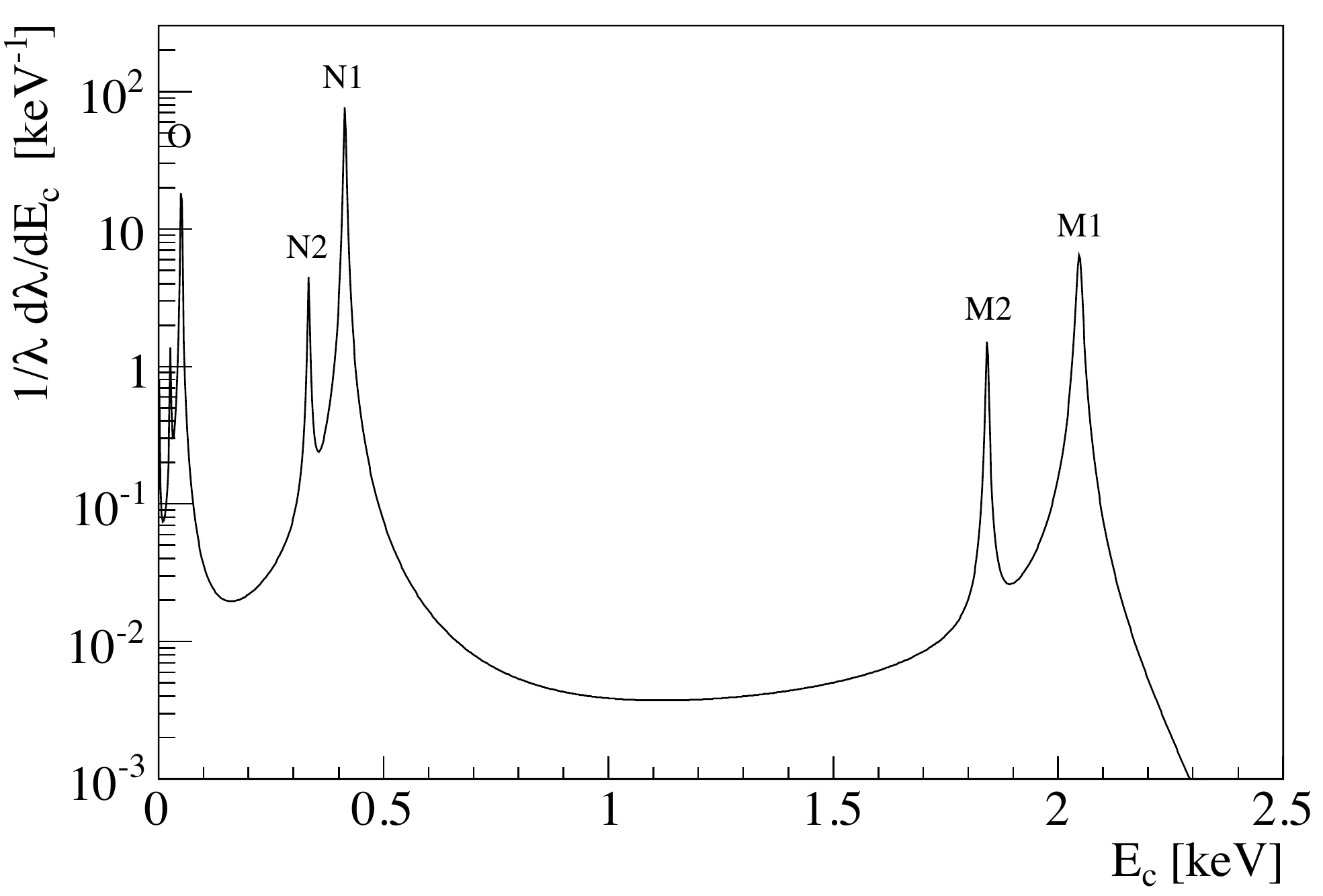}
 \caption{ Expected de--excitation energy spectrum of the  EC decay of $^{163}$Ho with $Q=2.5$~keV.  Detector resolution effects are not included. The parameters used in the calculation are discussed in Section \ref{numbers}.}
 \label{fig:fig1}
\end{figure}

\begin{figure}[htb]
 \centering
  \includegraphics[width=0.7\textwidth ]{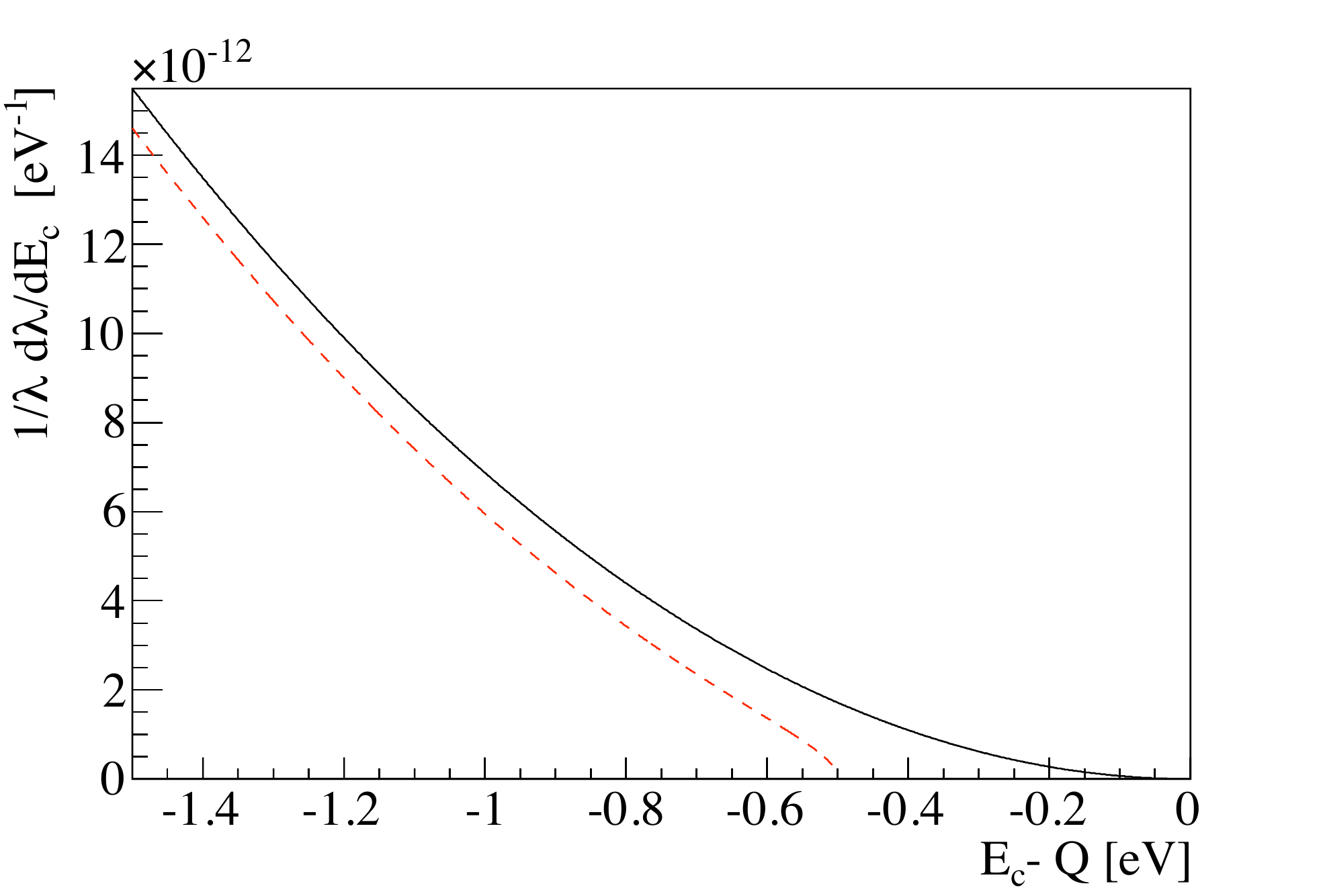}
 \caption{ Energy spectrum in EC decays for $^{163}$Ho with $Q=2.5$~keV near to the end point  
 for neutrino of zero mass (solid black line) or with  $m_{\nu}=0.5$~eV (dashed red line).}
 \label{fig:fig2}
\end{figure}

A calculated de--excitation energy spectrum is presented in fig.{\ref {fig:fig1}} and the effect of a nonzero neutrino mass near to the end point is shown in fig.{\ref {fig:fig2}}.

Bolometers, however, have the disadvantage of being slow, and therefore pile--up could be a problem. The way to tame it is to use a large number of smallish detectors \cite{Gatti:1997}.

\section{Relic antineutrino capture in $^{163}$Ho}
\label{theory2}
Consider now the capture by the original nucleus of a very low energy $\bar \nu_e$ and an electron from the $i$--th atomic shell:
\begin{equation}
\bar \nu_e + \, ^{163}{\rm Ho}  \to  \; ^{163}{\rm Dy}_i^* \; .
\label{reaction}
\end{equation}  
The procedure to evaluate the rate for this process has been presented in ref.~\cite{Cocco:2009rh}, following the formalism for EC decays introduced in ref.~\cite{Bambynek:1977zz} and  can be written as 
\begin {equation}
\label{reac-rate}
\lambda_{\bar \nu} = n_{\bar \nu} {G_{\beta}^2 \over 2} \; \sum_i n_i \,C_i  \,
 \beta_i^2 B_i\,  \rho_i(E_{\bar \nu})
\end{equation}
where $n_{\bar \nu}$ is the number density of incoming antineutrinos, $E_{\bar \nu}$ is their energy ($\simeq m_{\nu}$ for C$\nu$B) and $ \rho_i(E_{\bar \nu})$ is the density of final states. Again, the nuclear shape factors can be factored out of the sum. 

The final states in reaction (\ref{reaction}) are unstable and the final value of the de--excitation energy must be $Q + m_{\nu}$ for a zero energy incoming antineutrino. Even if this value  does not coincide with the maximum of the Breit--Wigner curve it can be reached anyhow, although of course the rate will be suppressed. As a consequence, the number of available final states per unit energy $\rho_i(E_{\bar \nu})$ defined in ref.~\cite{Cocco:2009rh} should be modified as follows:
 \begin{equation}
 \label{BW}
 \rho_i(E_{\bar \nu}) = \delta (E_{\bar \nu} + Q - E_i) \longrightarrow
 {1 \over \pi} \cdot {\Gamma_i/2 \over {(E_{\bar \nu}+Q-E_i)^2+\Gamma_i^2/4}} \;.
\end{equation}  
 The ratio of the rates of the two processes considered is approximately:
\begin {equation}
\label{ratio}
{\lambda_{\bar \nu} \over \lambda_{EC}} \simeq 2\,\pi^2\,n_{\bar \nu}\;{\sum_i n_i   \beta_i^2 B_i \;
  \rho_i(E_{\bar \nu})  \over \sum_i n_i  \beta_i^2 B_i (Q-E_i)^2} \;.
\end{equation}

As we will show in the next section, for C$\nu$B antineutrinos this is a {\it really} small number. However, the EC events that can be a background are only those falling in a narrow energy interval before the end point, namely
$$Q-\Delta-m_{\nu} \leq E_c \leq Q-m_{\nu}\;. $$
The fraction of EC events falling in this region is given by the so--called factor of merit:
\begin{equation}
\label{fig_of_merit}
F(\Delta, m_{\nu},Q) = {1 \over  \lambda_{EC}} \; \int_{Q-\Delta-m_{\nu}}^{Q-m_{\nu}}
{d \lambda_{EC}\over dE_c} dE_c\;.
\end{equation}
Neglecting the variations of the different Breit-Wigner over the  small scale $\Delta$ and neglecting the neutrino mass with respect to $Q-E_i$, one has:
\begin{align}
\label{fig1_of_merit}
F(\Delta, m_{\nu},Q)& \simeq {\Delta^3 \over 3} \left(1+{2 m_{\nu} \over \Delta} \right)^{3/2} \times\nonumber \\
&{ \sum_i n_i   \beta_i^2 B_i \; (\Gamma_i / 2\,\pi)\;\left[(Q-E_i)^2+\Gamma_i^2/4\right]^{-1}
\over \sum_i n_i   \beta_i^2 B_i\; (Q-E_i)^2} \; .
\end{align}

As a consequence, the ratio of the counting rates of the antineutrino capture and the EC decays near to the end point is:
\begin{eqnarray}
\label{final}
R(\Delta, m_{\nu}, Q) &=& {1 \over F(\Delta, m_{\nu},Q)}  {\lambda_{\bar \nu} \over \lambda_{EC}} \nonumber\\
&\simeq& 6\,\pi^2\, {n_{\bar \nu} \over \Delta^3} \left(1+ {2 m_{\nu} \over \Delta} \right)^{-{3 \over 2}}\;.
\end{eqnarray}
The above expression of $R(\Delta, m_{\nu}, Q)$ does not depend on $Q$ and is equal to the analogous result for $\beta$ decay \cite{Cocco:2007za}, showing that both types of radioactive decaying nuclei are in principle equally good as targets to detect C$\nu$B.

\section{Numerical results}
\label{numbers}
 We proceed to give numerical results, based on estimates found in the literature for the  parameters appearing in the previous equations. The levels of the electrons that can be captured are fully occupied ($n_i=1$). Their  binding energies and widths are reported in Table \ref{En-Wid}. 
Note that the real  values may slightly differ \cite{Bennett3} from these, obtained in dysprosium excitation, but they will be precisely determined in future calorimetric experiments, from  widths and positions of the peaks in the measured $E_c$ distribution, see fig.(\ref{fig:fig1}). 

\begin{table}[h]
\centering
\caption{Energy levels of the captured electrons, with their widths, for $^{163}$Dy \cite{param1}.}
\begin{tabular}{ccc}
\hline
Level&$E_i$ (eV)&$\Gamma_i$ (eV)\\
\hline
M$_1$ & 2047 & 13.2 \\
M$_2$ & 1842 & 6.0 \\
N$_1$ & 414.2 & 5.4  \\
N$_2$ & 333.5 & 5.3  \\
O$_1$ & 49.9 &   \\
O$_2$ & 26.3 &   \\
\hline
\end{tabular}
\label{En-Wid}
\end{table}

In Table \ref{waveratio}  we report the relative values of the squared wave functions, namely the ratios of the parameters $\beta_i^2/\beta_{\rm M_1}^2$. The exchange and overlap 
corrections\footnote{They are not given for all the levels needed in \protect\cite{Bambynek:1977zz}. Those given are less than $\sim 10\%$.} are neglected (i.e. $B_i\sim 1$). The validity of this approximation far from the peaks  may be doubted, however the {\it shape} of the spectrum in an interval of O($\Delta$) near to the end--point is anyhow determined by the neutrino phase-space factor. The {\it rate} at the end--point can be modified: a phenomenological model has been proposed in \cite{Riisager:1988wy}, where it was suggested that $F(\Delta, m_{\nu},Q)$ may be suppressed by a factor about 2. In this case, it is obvious that also the capture rate of C$\nu$B would be suppressed by the same amount, leaving the ratio $R(\Delta, m_{\nu}, Q)$ unchanged. In the following, we will derive results using our expressions, but keep in mind the possibility of a small further suppression in the counting rate. On the other hand, we are neglecting the overdensity due to gravitational clustering, that will certainly increase the rate. 

\begin{table}[h]
\centering
\caption{Electrons squared wave functions at the origin $\beta_i^2$ relative to $\beta_{\rm M_1}^2$ \cite{Band:1985gm}.}
\begin{tabular}{cc}
\hline
Levels&Ratio\\
\hline
M$_2$/M$_1$ & 0.0526 \\
N$_1$/M$_1$ & 0.2329 \\
N$_2$/M$_1$ & 0.0119  \\
O$_1$/M$_1$ & 0.0345 \\
O$_2$/M$_1$ & 0.0015 \\
P$_1$/M$_1$ & 0.0021  \\
\hline
\end{tabular}
\label{waveratio}
\end{table}

Assuming for the unknown (and not very relevant) parameters the values $\Gamma_i = (3,3,1)$~eV for the levels (O$_1$, O$_2$, P$_1$) and $E_i \sim 0$ for P$_1$, we find that the ratio of C$\nu$B antineutrino captures to the total EC events defined in eq.(\ref{ratio}) is:
\begin {equation}
\label{ratioval}
{\lambda_{\bar \nu} \over \lambda_{EC}} = (7.7\cdot10^{-22}, 5.8\cdot10^{-23}, 1.4\cdot10^{-23})
\end{equation}
for $Q=(2.3, 2.5, 2.8)~{\rm keV}$, values higher than the analogous result for tritium $\beta$--decays \cite{Cocco:2007za}:  $\lambda_{\nu} / \lambda_{\beta} = 6.6\cdot10^{-24}$. Assuming as an example a value of 0.5~eV for the neutrino mass and  $\Delta = 0.2$~eV, we have
\begin {equation}
\label{figmervall}
F(0.2~{\rm eV}, 0.5~{\rm eV},Q) =
 (3.6\cdot10^{-12}, 2.7\cdot10^{-13}, 6.5\cdot10^{-14})
\end{equation}
for $Q=(2.3, 2.5, 2.8)~{\rm keV}$, to be compared with the value $3\cdot10^{-14}$ for tritium. The half--lives are $T_{1/2} = 4570\;(12.32)$~y for $^{163}$Ho ($^3$H). Therefore we confirm that a calorimetric experiment with 
$^{163}$Ho, having a higher factor of merit, may be a competitor of $^3$H for hunting the neutrino mass effect. 

For the detection of C$\nu$B the ratio of the counting rates of the antineutrino capture and the EC decays near to the end point, $R(\Delta, m_{\nu}, Q)$, is equal to the corresponding quantity for $\beta$--decaying nuclei. We made an analysis including the effect of the detector resolution to determine the discovery potential of a future experiment using a $^{163}$Ho target. Let us consider the total number of signal events:
\begin{equation}
 S = {\lambda_{\bar \nu} \over \lambda_{EC}} {\log 2 \over T_{1/2}} N_A n_{\rm mol} t\;,
\end{equation}
 where $N_A$ is Avogadro's number, $n_{\rm mol}$ the number of mols, $t$ the exposure time, $T_{1/2}$ the half--life of $^{163}$Ho and assume that we require a minimum number of 10 events observed. Using the values in eq.(\ref{ratioval}) this correspond to a minimum quantity of $^{163}$Ho of (23.2, 307, 1274)~kg$\cdot$y for $Q = (2.3, 2.5, 2.8)$~keV. 
 
\begin{figure}[ht]
 \centering
  \includegraphics[width=0.7\textwidth ]{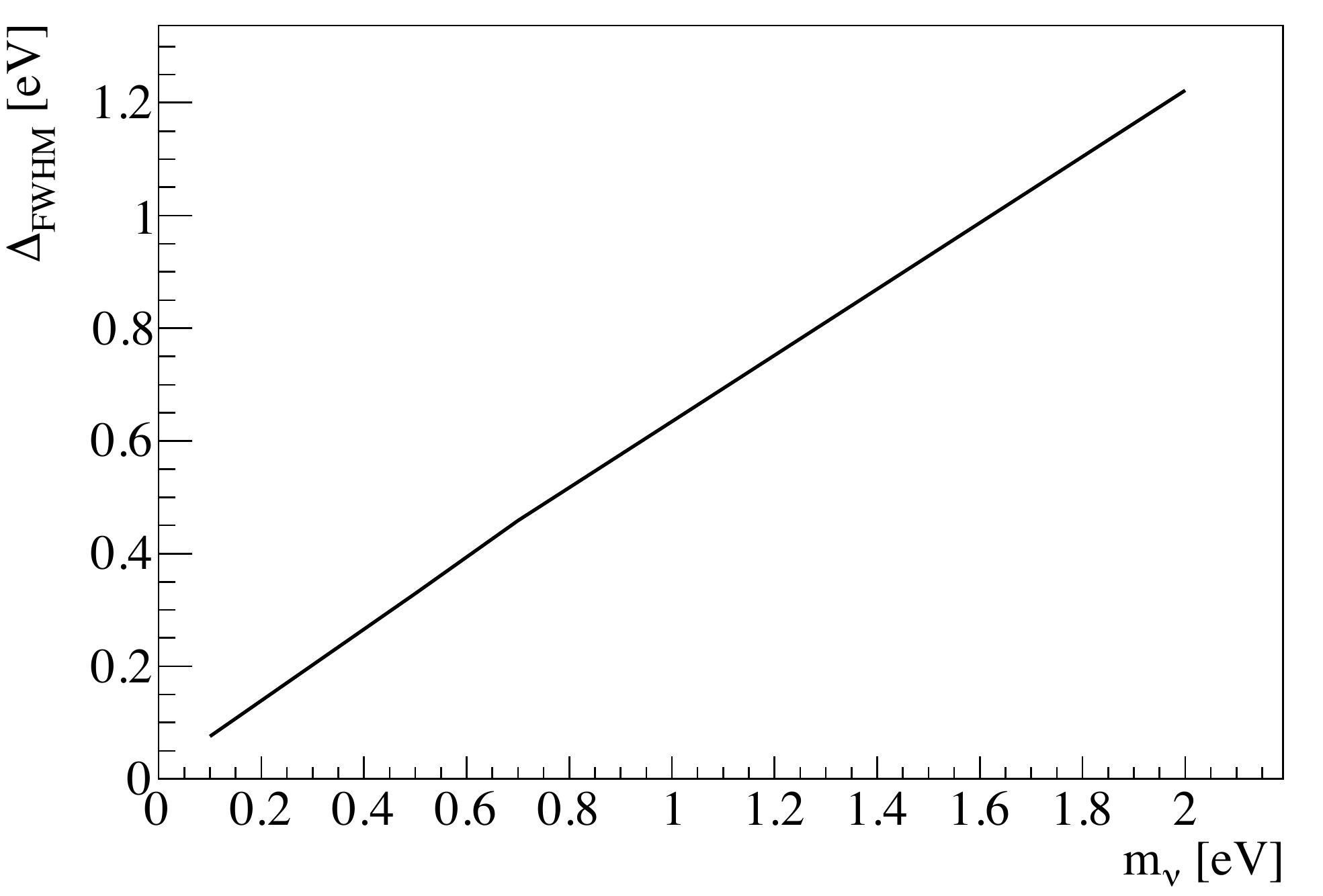}
 \caption{Detector resolution needed as a function of the neutrino mass. The discovery region, where $S / \sqrt {B} \geq 5$ for $S=10$, falls below the line.}
 \label{fig:fig3}
\end{figure}

 The number of background events falling in an interval of amplitude 
 $ \Delta_{\rm FWHM} = 2.35\, \Delta$ centered at $Q+m_{\nu}$ can be obtained by convoluting the energy distribution in EC events with a gaussian of variance $\Delta^2$. Defining
$$ 
b(\Delta, m_{\nu},Q) = {1 \over \lambda_{EC}}\,{1 \over \sqrt{2 \pi} \Delta} \; \int_{Q+m_{\nu}-\Delta_{\rm FWHM}/ 2}^{Q+m_{\nu}+\Delta_{\rm FWHM}/ 2} dE' \int_{0}^{Q-m_{\nu}} dE 
{d \lambda_{EC}\over dE}(E) \; e^{-{(E-E')^2 \over 2 \Delta^2}} $$
the number of background events is:
 \begin{equation}
\label{bckg1}
B(\Delta, m_{\nu},Q) = b(\Delta, m_{\nu},Q) \;{\log 2 \over T_{1/2}} N_A n_{\rm mol} t \;.
\end{equation}
Defining the statistical significance as $S / \sqrt {B}$,  in fig.{\ref {fig:fig3}}
 we present the boundary of the discovery region, where the statistical significance  is larger than 5,  in the plane ($m_{\nu}$, $\Delta_{\rm FWHM}$). The dependence on $\Delta$ is so sharp that the variation with $Q$ cannot be appreciated given the thickness of the line.

The boundary of the discovery region presented in fig.{\ref {fig:fig3}} also applies to an experiment using a tritium target (and looking for neutrinos in the C$\nu$B instead of antineutrinos). Due to the different half--life and mass number, in this case  to have a minimum of 10 signal events one needs 137~g$\cdot$y of $^3$H. At present, the requirements for both EC and $\beta$ decaying nuclei seem very demanding and we do not know which of the two very different technologies may have more chances. Note that our estimates are pessimistic, since the inclusion of gravitational clustering effects would enhance the number of signal events. 

 \section{Conclusions}
  We have presented in this work an estimate of the requirements for an experiment aiming to detect antineutrinos in the C$\nu$B using a target of $^{163}$Ho.  The request to have a reasonable number of events in the signal 
 gives a constraint on the mass of the source and on the exposure time that is very sensitive to the $Q$--value, not yet well known. Assuming $Q=2.5$~keV one would need ten years of observation of a source of 30 kilograms to have 10 events of signal. Even more stringent maybe are the sensitivity requirements, that instead are practically independent on $Q$: for a neutrino mass of 0.5~eV, for instance, one would need a resolution FWHM better than 0.33~eV in order to attain a statistical significance of 5, the usual requirement for a discovery.
 Nonetheless, maybe the neutrino mass is higher, the $Q$--value is smaller and the experimental ingenuity may arrive at resolutions better than the present ones.

\end {document}